
%
%
\input phyzzx
\nopubblock
\pubnum={91-53}
\date={November 1991}
\titlepage
\title{$N=2$ String as a Topological Conformal Theory}
\author{Joaquim Gomis\footnote{\ast}{Permanent address: Departament
 d'Estructura i Constituents de la Mat\`eria, Universitat de Barcelona,
 Diagonal 647, E-08028 Barcelona, Catalonia, Spain. Electronic address:
 quim@ebubecm1.bitnet}and Hiroshi Suzuki\footnote{\dag}{JSPS Junior
 Scientist Fellow. Also at Department of Physics, Hiroshima University,
 Higashi-Hiroshima 724, Japan. Electronic address:
 suzuki@jpnrifp.bitnet}}
\address{
Uji Research Center, Yukawa Institute for Theoretical Physics,
Kyoto University, Uji 611, Japan}
\abstract{We prove that critical and subcritical $N=2$ string theory
 gives a realization of an $N=2$ superfield extension of the
 topological conformal algebra. The essential observation is the
 vanishing of the background ghost charge.}
%
\sequentialequations 
%
%
\def\journal#1&#2(#3){\unskip, \sl #1~\bf #2 \rm (19#3) }
\def\journals#1&#2(#3){\unskip; \sl #1~\bf #2 \rm (19#3) }
\REF\ADE{
 M.\ Ademollo, L.\ Brink, A.\ D'Adda, R.\ D'Auria, E.\ Napolitano,
 S.\ Sciuto, E.\ Del Giudice, P.\ Di Vecchia, S.\ Ferrara, F.\ Gliozzi,
 R.\ Musto and R.\ Pettorino
    \journal Phys.\ Lett.&B62(76)105;\hfil\break
 M.\ Ademollo, L.\ Brink, A.\ D'Adda, R.\ D'Auria, E.\ Napolitano,
 S.\ Sciuto, E.\ Del Giudice, P.\ Di Vecchia, S.\ Ferrara, F.\ Gliozzi,
 R.\ Musto, R.\ Pettorino and J.\ H. Schwarz
    \journal Nucl.\ Phys.&B111(76)77.}
\REF\BRU{
 D.\ J.\ Bruth, D.\ B.\ Fairle and R.\ G.\ Yates
    \journal Nucl.\ Phys.&108(76)310;\hfil\break
 L.\ Brink and J.\ H.\ Schwarz
    \journal Nucl.\ Phys.&B121(77)285;\hfil\break
 E.\ S.\ Fradkin and A.\ A.\ Tseytlin
    \journal Phys.\ Lett.&B106(81)63;\hfil\break
 P.\ Bouwknegt and P.\ van Nieuwenhuizen
    \journal Class.\ Quantum Grav.&3(86)207;\hfil\break
 A.\ D'Adda and F.\ Lizzi
    \journal Phys.\ Lett.&B191(87)85;\hfil\break
 S.\ D.\ Mathur ans S.\ Mukhi
    \journal Nucl.\ Phys.&B302(88)130;\hfil\break
 N.\ Ohta and S.\ Osabe
    \journal Phys.\ Rev.&D39(89)1641;\hfil\break
 M.\ Corvi, V.\ A.\ Kosteleck\'y and P.\ Moxhay
    \journal Phys.\ Rev.&D39(89)1611.}
\REF\BIL{
 A.\ Bilal
    \journal Phys.\ Lett.&B180(86)255;\hfil\break
 A.\ R.\ Bogojevic and Z.\ Hlousek
    \journal Phys.\ Lett.&B179(86)69;\hfil\break
 S.\ D.\ Mathur ans S.\ Mukhi
    \journal Phys.\ Rev.&D36(87)465.}
\REF\GOM{
 J.\ Gomis
    \journal Phys.\ Rev.&D40(89)408.}
\REF\OOG{
 H.\ Ooguri and C.\ Vafa
    \journal Nucl.\ Phys.&B361(91)469.}
\REF\WITT{
 E.\ Witten
    \journal Commun.\ Math.\ Phys.&117(88)353.}
\REF\WIT{
 E.\ Witten
    \journal Commun.\ Math.\ Phys.&118(88)411
    \journals Nucl.\ Phys.&B340(90)281.}
\REF\EGU{
 T.~Eguchi and S.-K.~Yang
    \journal Mod.~Phys.~Lett.&A5(90)1693.}
\REF\FUJ{
 K.\ Fujikawa
    \journal Phys.\ Rev.&D25(82)2584.}
\REF\FRI{
 D.\ Friedan, E.\ Martinec and S.\ Shenker
    \journal Nucl.\ Phys.&B271(86)93.}
\REF\MAR{
 S.\ P.\ Martin
    \journal Phys.\ Lett.&B191(87)81.}
\REF\NOJ{
 S.\ Nojiri
    \journal Phys.\ Lett.&B262(91)419
    \journals Phys.\ Lett.&B264(91)57.}
\REF\DIS{
 J.\ Distler, Z.\ Hlousek and H.\ Kawai
    \journal Int.\ J.\ Mod.\ Phys.&A5(90)391;\hfil\break
 I.\ Antoniadis, C.\ Bachas and C.\ Kounnas
    \journal Phys.\ Lett.&B242(90)185.}
\REF\YOU{
 K.\ Fujikawa, T.\ Inagaki and H.\ Suzuki
    \journal Phys.\ Lett.&B213(88)279
    \journals Nucl.\ Phys.&B332(90)499.}
\REF\OUR{
 J.\ Gomis and H.\ Suzuki, to appear.}
\REF\FUJI{
 K.\ Fujikawa and H.\ Suzuki
    \journal Nucl.\ Phys.&B361(91)539.}
\REF\LI{
 K.\ Li
    \journal Nucl.\ Phys.&B346(90)329.}
\REF\ROC{
 J.\ Gomis and J.\ Roca
    \journal Phys.\ Lett.&B268(91)197.}
\REF\NEL{
 S.\ Govindarajan, P.\ Nelson and S.-J.\ Rey
    \journal Nucl.\ Phys.&B365(91)633.}
The $N=2$ superstring introduced by Ademollo {\it et.\ al.}
 [\ADE,\BRU,\BIL,\GOM,\OOG] has the critical dimension $2$, there
 are no transverse  degrees of freedom, the physical spectrum contains
 a finite number of particles, all massless and bosonic. There is a
 general belief that this is topological quantum theory. In this note
 we will prove that critical and subcritical $N=2$ strings are a
 topological field theories [\WITT,\WIT] in the sense that the
 reparametrization BRST current algebra gives a realization of an
 $N=2$ superfield extension of the topological conformal algebra
 [\EGU,\WITT]. The key ingredient for this proof is that $N=2$ string
 has no total background ghost charge and therefore no ghost number
 anomaly [\FUJ].

Let us first recall some basic facts of $N=2$ string in superfield
 formalism\rlap,\footnote{\ast}{We follow the notation of
 ref.\ [\GOM].}in $N=2$ superspace is described in terms of bosonic
 $(z,\bar z)$ and fermionic $(\theta^\pm,\bar\theta^\pm)$ coordinates.
 We can define covariant derivatives
$$
 D^\pm={\partial\over{\partial\theta^\mp}}+\theta^\pm\partial,\quad
 \bar D^\pm={\partial\over{\partial\bar\theta^\mp}}
  +\bar\theta^\pm\bar\partial.
\eqno(1)
$$
The action can be written in terms of two superfields
 $S^\mu(z,\bar z,\theta^+,\bar\theta^+,\theta^-,\bar\theta^-)$ and
 \break
$S^{\mu\ast}(z,\bar z,\theta^+,\bar\theta^+,\theta^-,\bar\theta^-)$
 satisfying two constraints
$$
 D^-S^\mu=\bar D^-S^\mu=0
\eqno(2)
$$
and
$$
 D^+S^{\mu\ast}=\bar D^+S^{\mu\ast}=0.
\eqno(3)
$$
The action is [\ADE]
$$
 A=\int dzd\bar z\int d\theta^+d\bar\theta^+d\theta^-d\bar\theta^-
  S^{\mu\ast} S^\mu.
\eqno(4)
$$
The solution of the equation of motions
$$
 D^+\bar D^+S^\mu=0,\quad\bar D^-D^-S^{\mu\ast}=0
\eqno(5)
$$
can be written as
$$
 S^\mu=S_1^\mu+S_2^\mu
\eqno(6)
$$
where
$$
\eqalign{
 &D^-S_1^\mu=\bar D^-S_1^\mu=\bar D^+S_1^\mu=0,
\cr
 &D^-S_2^\mu=\bar D^-S_2^\mu=D^+S_2^\mu=0.
\cr
}
\eqno(7)
$$
A real superfield $X^\mu$ is constructed via
$$
 X^\mu\left(z,\theta^+,\theta^-\right)
  =S_1^\mu\left(z+\theta^-\theta^+,\theta^-\right)
  +S_1^{\mu\ast}\left(z+\theta^+\theta^-,\theta^+\right).
\eqno(8)
$$
The components of $X^\mu(Z)$ are
$$
 X^\mu(Z)=X^\mu(z)+\theta^-\psi^{+\mu}(z)+\theta^+\psi^{-\mu}(z)
  +i\theta^-\theta^+\partial Y^\mu(z)
\eqno(9)
$$
where $X^\mu(z)$ and $Y^\mu(z)$ are free bosonic fields and
 $\psi^{\pm\mu}(z)$ are free fermions\rlap.\footnote{\ast}{We consider
 only the holomorphic part.}

We have the following operator product expansion
$$
 X^\mu(Z_a)X^\nu(Z_b)\sim\eta^{\mu\nu}\ln Z_{ab}
\eqno(10)
$$
where $Z_{ab}$ is
$$
 Z_{ab}=z_a-z_b-\left(\theta_a^+\theta_b^-+\theta_a^-\theta_b^+\right).
\eqno(11)
$$
$N=2$ primary conformal superfields $\psi_q^h(Z)$ are characterized by
 a weight $h$ and a charge $q$. They have the following OPE with the
 energy momentum tensor $T(Z)$
$$
\eqalign{
 T(Z_a)\psi_q^h(Z_b)&\sim
 h\,{{\theta_{ab}^-\theta_{ab}^+}\over{Z_{ab}^2}}\,\psi_q^h(Z_b)
 -{q\over{2Z_{ab}}}\psi_q^h(Z_b)
\cr
 &\quad+{1\over{2Z_{ab}}}\left(\theta_{ab}^-D_b^+
 -\theta_{ab}^+D_b^-\right)
 \psi_q^h(Z_b)
 +{{\theta_{ab}^-\theta_{ab}^+}\over{Z_{ab}}}
 \partial_{z_b}\psi_q^h(Z_b)
\cr
}
\eqno(12)
$$
where $\theta_{ab}^\pm=\theta_a^\pm-\theta_b^\pm$.
 The contribution to the energy momentum tensor from $X^\mu$ is
$$
T^X(Z)={1\over2}D^-X^\mu D^+X^\mu(Z).
\eqno(13)
$$

$N=2$ superstring action is invariant under several local gauge
 transformations. We will work in the superconformal gauge. Gauge
 fixing generates a Faddeev--Popov determinant expressible as a
 superfield action using $N=2$ superfield ghost $C$ and antighost $B$
$$
\eqalign{
 &C\equiv c+i\theta^+\gamma^--i\theta^-\gamma^
  ++i\theta^-\theta^+\xi,
\cr
 &B\equiv-i\eta-i\theta^+\beta^--i\theta^-\beta^+
 +\theta^-\theta^+b.
\cr
}
\eqno(14)
$$
The ghosts $c$ and $b$ are for the $\tau$-$\sigma$ reparametrization
 invariances, $\gamma^\pm$ and $\beta^\pm$ are the super ghosts for
 the two local supersymmetry transformations and $\xi$ and $\eta$ are
 the ghosts associated with the local U(1) symmetry. Their Lagrangians
 are the first order systems with background charge $Q$ [\FRI] and
 statistics $\epsilon$ of $(Q,\epsilon)=(-3,+)$, $(2,-)$ and $(-1,+)$
 respectively. Notice that the total background ghost charge vanishes.
 The ghost action in terms of superfield is given by
$$
 A_{\rm gh}={1\over\pi}\int d^2zd\theta^+d\theta^-B\bar\partial\,C
 +({\rm c.\ c.}).
\eqno(15)
$$
The non-zero fundamental operator product expansion for the
 holomorphic part is
$$
 C(Z_a)B(Z_b)\sim
 {{\theta_{ab}^-\theta_{ab}^+}\over{Z_{ab}}}
 \sim B(Z_a)C(Z_b).
\eqno(16)
$$
The ghost energy momentum tensor is
$$
 T^{\rm gh}(Z)=\partial(CB)(Z)-{1\over2}D^+CD^-B(Z)
 -{1\over2}D^-CD^+B(Z).
\eqno(17)
$$
The superfields $B(Z)$ and $C(Z)$ are $q=0$ conformal fields with
 $h=+1$ and $h=-1$.

If we consider the total energy momentum tensor
$$
 T=T^X+T^{\rm gh}
\eqno(18)
$$
the OPE of $T$ with itself becomes
$$
\eqalign{
 T(Z_a)T(Z_b)&\sim
 {{D-2}\over{4Z_{ab}^2}}
 +{{\theta_{ab}^-\theta_{ab}^+}\over{Z_{ab}^2}}\,T(Z_b)
 +{1\over{2Z_{ab}}}\left(\theta_{ab}^-D_b^+-\theta_{ab}^+D_b^-\right)
  T(Z_b)
\cr
 &\quad+{{\theta_{ab}^-\theta_{ab}^+}\over{Z_{ab}}}
 \,\partial\,T(Z_b)
\cr
}
\eqno(19)
$$
where $D$ is the dimension of the target space. For
 $D=2$\rlap,\footnote{\ast}{A first attempt to interpret this theory
 as a four dimensional $(2,2)$ theory was done by D'Adda and Lizzi in
 ref.\ [\BRU].}$T$ is a $q=0$, $h=0$ conformal superfield.

The $N=2$ BRST charge $Q_B$ [\BIL] is
$$
 Q_B=\oint dZ\,C\left(T^X+{1\over2}T^{\rm gh}\right)(Z)
\eqno(20)
$$
where
$$
 \oint dZ=\oint{{dz}\over{2\pi i}}\int d\theta^+d\theta^-
\eqno(21)
$$
and when $D=2$ it satisfies
$$
 Q_B^2=0
\eqno(22)
$$
since the total energy momentum tensor has no anomalies. Furthermore
 one can check that
$$
 T(Z)=\left\{Q_B,B(Z)\right\}.
\eqno(23)
$$
Now we can construct the ghost number current $j_{\rm ghost}$
$$
 j_{\rm ghost}(Z)=-BC(Z).
\eqno(24)
$$
Since the total background ghost charge is zero and this current is not
 anomalous, the ghost current is a primary field with $q=0$ and $h=0$.
 The previous facts are characteristic of $N=2$ string, for $N=0$ and
 $N=1$ due to the non-vanishing of the background ghost charge, the
 ghost current is anomalous and the ghost current is not a primary
 field [\FRI,\MAR].

Now let us construct the BRST current $j_B$, we will use a relation
$$
 j_B(Z)=-\left\{Q_B,j_{\rm ghost}(Z)\right\}
\eqno(25)
$$
in such a way that $j_B$ will be BRST invariant. Explicitly one finds
$$
\eqalign{
 j_B(Z)&=C(Z)\left(T^X+{1\over2}T^{\rm gh}\right)
\cr
 &\quad+{1\over4}D^-\left[C\left(D^+C\right)B\right]
       +{1\over4}D^+\left[C\left(D^-C\right)B\right]
\cr
}
\eqno(26)
$$
where the total derivative pieces\footnote{\ast}{The sign of the last
 term in eq.\ (26) is different from the expression of the BRST
 current in ref.\ [\GOM], which is primary however not BRST
 invariant.}ensure that $j_B(Z)$ is a primary superfield with $q=0$
 and $h=0$.

At this point we can define the $N=2$ superfield extension of the
 topological conformal algebra. The generators are
$$
\eqalign{
 &T(Z)\equiv T(Z),
\cr
 &G(Z)\equiv j_B(Z),
\cr
 &\bar G(Z)\equiv B(Z),
\cr
 &J(Z)\equiv j_{\rm ghost}(Z).
\cr
}
\eqno(27)
$$
The relevant operator product expansions are
$$
\eqalign{
 T(Z_a)\Psi(Z_b)&\sim
 h\,{{\theta_{ab}^-\theta_{ab}^+}\over{Z_{ab}^2}}\,\Psi(Z_b)
 +{1\over{2Z_{ab}}}\left(\theta_{ab}^-D_b^+-\theta_{ab}^+D_b^-\right)
 \Psi(Z_b)
\cr
 &\quad+{{\theta_{ab}^-\theta_{ab}^+}\over{Z_{ab}}}
 \,\partial_{z_b}\Psi(Z_b)
\cr
}
\eqno(28)
$$
for $\Psi=T$, $G$, $\bar G$, and $J$ with $h=1$, $0$, $1$, and $0$
 respectively and
$$
\eqalign{
 &G(Z_a)\bar G(Z_b)\sim
 {1\over{2Z_{ab}}}\left(\theta_{ab}^-D_b^+-\theta_{ab}^+D_b^-\right)
 J(Z_b)
 +{{\theta_{ab}^-\theta_{ab}^+}\over{Z_{ab}}}
 \,T(Z_b),
\cr
 &G(Z_a)G(Z_b)\sim0,
\cr
 &\bar G(Z_a)\bar G(Z_b)\sim0,
\cr
 &J(Z_a)J(Z_b)\sim0,
\cr
 &J(Z_a)G(Z_b)\sim
  {{\theta_{ab}^-\theta_{ab}^+}\over{Z_{ab}}}\,G(Z_b),
\cr
 &J(Z_a)\bar G(Z_b)\sim
  -{{\theta_{ab}^-\theta_{ab}^+}\over{Z_{ab}}}\,\bar G(Z_b).
\cr
}
\eqno(29)
$$

Summing up we should conclude that the critical $N=2$ string is a
 topological field theory in the sense that the reparametrization BRST
 current algebra gives a representation of a $N=2$ superfield
 extension of the topological conformal
 algebra\rlap.\footnote{\ast}{Our algebra in eqs.\ (28) and (29),
 in terms of the component fields, is not same as the twisted $N=4$
 superconformal algebra by Nojiri [\NOJ].}Actually Ooguri and Vafa
 [\OOG] have shown an close relation between $N=2$ string and self-dual
 gravity and self-dual Yang--Mills, it will be interesting to study
 the implications of our results from this type of theories.

With respect to $N=2$ subcritical strings or $N=2$ super Liouville
 theory we can also show that these theories are topological. For
 analyzing these theories one might do some hypothesis about a local
 ansatz for the Jacobian between a non-trivial measure and the free
 measure [\DIS]. Or we might construct an effective theory [\YOU] based
 on anomalous identities associated to the superconformal, the BRST
 and the ghost number symmetries. In the latter, the renormalization
 of the coupling constant (see below) is considered as the one-loop
 order effect of the BRST invariant measure. We will follow the second
 procedure with the vanishing cosmological constant, generalizing the
 results of the $N=0$ and $N=1$ case [\YOU] to $N=2$ super Liouville
 theory. To construct the relevant operators for the effective theory
 [\OUR] we should introduce a Liouville superfield
$$
 \Phi(Z)=\phi(z)+\theta^-\phi^+(z)+\theta^+\phi^-(z)
  +i\theta^-\theta^+\partial\rho(z)
\eqno(30)
$$
with the operator product
$$
 \Phi(Z_a)\Phi(Z_b)=\ln Z_{ab}.
\eqno(31)
$$
The form of the energy momentum tensor is
$$
 T=T^X+T^{\rm gh}+T^{\rm Liouville}
\eqno(32)
$$
where $T^X$ and $T^{\rm gh}$ are given by eqs.\ (13) and (17). To
 construct $T^{\rm Liouville}$ we make an observation that apart from
 an inhomogenious piece, Liouville superfield behaves as $X^\mu$. Then
 we have
$$
 T^{\rm Liouville}={1\over2}D^-\Phi D^+\Phi+\kappa\partial\Phi
\eqno(33)
$$
where $\kappa$ is the renormalized coupling constant of the effective
 theory and algebraically it can be calculated imposing that the total
 energy momentum tensor $T$ behaves as a superfield with $q=0$ and
 $h=0$. This condition implies
$$
 \kappa^2={{1-D}\over4}.
\eqno(34)
$$

For the ghost number current as we have described before there is no
 ghost number anomaly, thus the current is given by
$$
 j_{\rm ghost}(Z)=-BC(Z)
\eqno(35)
$$
as for the critical string case, $j_{\rm ghost}$ is a superfield with
 $q=0$ and $h=0$.

The BRST charge associated to the BRST symmetry is given by
$$
 Q_B=\int DZ\,C\left(T^X+T^{\rm Liouville}+{1\over2}T^{\rm gh}\right)
\eqno(36)
$$
and it satisfies $Q_B^2=0$ for any $D$. Furthermore one can also check
$$
 T=\left\{Q_{\rm BRST},B\right\}.
\eqno(37)
$$
Then we determine the total divergent pieces of the BRST current by
 using eq.\ (25). Explicitly one finds
$$
\eqalign{
 j_B(Z)&=C(Z)\left(T^X+T^{\rm Liouville}+{1\over2}T^{\rm gh}\right)
\cr
 &\quad+{1\over4}D^-\left[C\left(D^+C\right)B\right]
       +{1\over4}D^+\left[C\left(D^-C\right)B\right]
\cr
}
\eqno(38)
$$
where the total derivative pieces ensures that $j_B(Z)$ is a primary
 $N=2$ superfield with $q=0$ and $h=0$.

At this point it is interesting to comment about the differences of
 the analogous calculation for the $N=0$ and $N=1$ cases [\YOU]. The
 two main differences with respect to those cases are i) the
 appearance of a Liouville term in the ghost number current ii) the
 appearance of a divergent terms with a Liouville field in the
 expression of BRST current. The presence of these terms implies that
 the previous theories are not topological for arbitrary values of the
 dimension [\FUJI]. The main reason for that is the anomalous behavior
 of the ghost number current.

As in the critical string case $T$, $j_B$, $j_{\rm gh}$ and $B$ give a
 representation of the $N=2$ superfield extension of the topological
 conformal algebra eqs.\ (28) and (29). It is also useful to see how
 the critical string can be considered as a subcritical string in
 dimension $1$ plus the Liouville superfield, in fact for this
 situation, $\kappa=0$ and all the operators of the effective theory
 coincide with the ones of the critical string, since in this case
 there is no restriction for the possible values of $D$. Distler,
 Hlousek and Kawai [\DIS] notice already that their local ansatz for
 the Jacobian works for every $D$. It will be interesting to find the
 dimension $D$ in which one can establish the equivalence with the
 recent proposed $N=2$ topological supergravity [\LI,\ROC,\NEL] with
 gauge group ${\rm Osp}(2|2)$.

In summary in this note we have shown that the $N=2$ string theory is
 a topological field theory in any dimension $D$ in the sense that the
 reparametrization BRST current algebra realizes an $N=2$ superfield
 extension of the topological conformal algebra. This result is
 peculiar to the $N=2$ string and it is due to the fact that the
 background charge vanishes and therefore there is no ghost number
 anomaly.
\ack
We acknowledge Prof.\ K.\ Fujikawa for discussions and a critical
 reading of the manuscript. We also thank Prof.\ M.\ Ninomiya for a
 reading of the manuscript. One of us (J.\ G.) would like to thank
 Yukawa Institute for Theoretical Physics for hospitality.

\refout
\bye